\documentstyle[12pt,world_sci]{article}
\begin{document}

\title{{\bf $Z$ POLARIZATION  IN $pp\rightarrow ZZ\rightarrow \ell^+\ell^-\nu
\bar{\nu}$ AT THE LHC}}
\author{M. J. DUNCAN and M. H. RENO\thanks{Presenter.
Work supported in part
by National Science Foundation Grant No. PHY-9307213.}\\
{\em Department of Physics and Astronomy, University of Iowa, Iowa City,\\
Iowa 52242, USA}}

\maketitle
\setlength{\baselineskip}{2.6ex}

\begin{center}
\parbox{13.0cm}
{\begin{center} ABSTRACT \end{center}
{\small \hspace*{0.3cm}
We evaluate the feasibility of measuring the $Z$ polarization
from a heavy Higgs signal in the decay channel $H\rightarrow ZZ
\rightarrow \ell^+
\ell^-\nu\bar{\nu}$. Including gluon fusion production of the Higgs,
continuum production of $Z$ pairs and the QCD background of single $Z$
production with a missing jet, we find that the average value of
a new variable to measure the $Z$ polarization, as a function of
transverse mass, will demonstrate the existence of a Higgs boson for
heavy Higgs masses  up to 800 GeV at the LHC with an integrated
luminosity of 10$^5$ pb$^{-1}$.
}}
\end{center}

\section{Introduction}

The `gold-plated' heavy Higgs decay mode $H\rightarrow Z Z\rightarrow
\ell_1^+\ell_1^-\ell_2^+\ell_2^-$ has received much attention in
the literature because of the opportunity for a fully reconstructed Higgs
signal.$^1$ %\cite{\goldplate}
For Higgs masses near the unitarity limit of $~800$ GeV,$^2$ %\cite{\unitarity}
however,
the event rates are low
for $pp$ production of the Higgs boson
at the Large Hadron Collider (LHC) center of mass energy $\sqrt{S}=14$ TeV.
To augment the Higgs search in the high mass region,
the Higgs decay mode $H\rightarrow ZZ\rightarrow \ell^+\ell^-\nu\bar{\nu}$
should also be considered as it
is enhanced relative to the gold-plated decay mode by a factor of
six.$^3$ %\cite{\cc}
This mode is easily identifiable by the large missing
transverse momentum in the event and a pair of charged leptons that
reconstruct to the $Z$ mass. In addition to searching for an enhancement
in the event rate, best identified in the transverse mass
distribution,$^4$ %\cite{\bhp}
one can also use polarization information to identify a Higgs signal.
The Higgs boson decays preferentially into longitudinally
polarized $Z$'s, ($Z_L$)
while the irreducible and reducible backgrounds to
$pp\rightarrow \ell^+\ell^-+$missing $p_T$ involve primarily transversely
polarized $Z$'s ($Z_T$).
Polarization methods have been adopted for the gold-plated
mode$^5$ %\cite{\malcolm}
and the $WW$ final state$^6$, %\cite{\ww}
and we have
suggested that this be applied to the case of $pp\rightarrow ZZ\rightarrow
\ell^+\ell^-\nu\bar{\nu}$.$^7$ %\cite{\dr}
Here, we demonstrate that even with
the QCD background included, the LHC with an integrated luminosity of
${\cal L}=10^5$ pb$^{-1}$ has a capability of distinguishing Higgs boson
production in the $\ell^+\ell^-\nu\bar{\nu}$ decay mode
from the background signal for a range of
heavy Higgs masses $m_H$ up to 800 GeV.
We consider here masses between 400 GeV and 800 GeV.

\section{Signal and Backgrounds}

The Higgs coupling to other particles is proportional to the particle's
mass, so the main contributions to Higgs production involve top quarks
and weak gauge bosons. Gunion {\it et al.}$^8$ have shown that
production of the Higgs, with $t\bar{t}$ in the final state is very small.
The main Higgs production processes are
gluon fusion into a top quark loop, which couples to the Higgs, and
vector boson fusion into a Higgs. For a top quark mass of 175 GeV,
the gluon fusion mechanism dominates vector boson fusion for the full
range of Higgs masses of interest: $m_H=400-800$ GeV.
The vector boson fusion cross section $qq\rightarrow qqZZ$
with an $s$-channel Higgs,
for Higgs masses between
400 GeV and 800 GeV, lies below the gluon fusion cross section
by a factor of $\sim 10$ at the low mass end, to a factor of $\sim 3$
at the high mass end.  Baur and Glover$^9$ have done the full calculation
of $ZZ$+2 parton production and have found that the $s$-channel approximation
overestimates the contribution of $qq\rightarrow qqZZ$ to a polarization
measurement because of additional $t$-channel and non-resonant contributions.
To be on the conservative side in our analysis, we include only the
gluon fusion mechanism.

Higgs decay into two $Z$'s goes primarily into longitudinal $Z$'s. It
is this effect which we exploit in our analysis of the signal.
The large irreducible background to
$pp\rightarrow H\rightarrow ZZ$ is continuum
production of $Z$ pairs via $q\bar{q}\rightarrow ZZ$.
The dominant contribution
to this background rate comes from production of a pair of transverse $Z$'s.

The main reducible background is from QCD corrections to single $Z$ production:
a $Z$ is produced at large transverse momentum, which decays into an
$\ell^+\ell^-$ pair, and is
accompanied by jets that are missed in the detector because
they go down the beam pipe. To model this background, we use the
matrix elements for $q\bar{q}\rightarrow Zg\rightarrow e^+e^-g$ and
crossed diagrams.
We count in the background rate only the part of the cross section
where the charged leptons have rapidity $|y_\ell |<y_\ell^c$, so that they are
in the central region, and a large final state parton rapidity $|y_p|>y_p^c$.

In Fig. 1, we show the transverse mass distribution for the three contributions
above. Here, the transverse mass $m_T$ is defined to be
$$m_T^2=[(\vec{p}_T\, ^2+m_Z^2)^{1/2}+(\vec{p\llap/}_T\, ^2+m_Z^2)^{1/2}]^2-
(\vec{p}_T
+\vec{p\llap/}_T)^2
$$
where $\vec{p}_T$ is the $Z$ boson transverse momentum reconstructed from
$\ell^+\ell^-$ momenta and $\vec{p\llap/}_T$
is the missing transverse momentum.
In our Monte Carlo, with these
three processes, all contributions
have balanced $p_T$: $\vec{p}_T=-\vec{p\llap/}_T$. The dashed line in the
figure indicates the gluon fusion production of a 500 GeV Higgs, the
solid line shows the continuum
production of $ZZ$ and the
dot-dashed line shows the rate from QCD production of $Z+$missing jet.
Here the rapidity cuts of $y_p^c=y_\ell^c=3$ are used, and in all the
figures, a charged lepton transverse momentum cut of $p_T>20$ GeV is applied.
The heavy solid line
is the total of the three contributions.

A longitudinal
polarization vector for a massive particle, subjected to an arbitrary
boost, will not, in general, retain its longitudinal character, so the
statement that the $Z_LZ_L$ production rate is enhanced
by a specific amount in Higgs production is a frame dependent statement.
For a sufficiently
heavy Higgs, the parton center of mass frame (the Higgs rest frame)
and the hadron center of mass
frame coincide.

\vspace*{5.0cm}
\begin{center}
{\small Fig. 1. The transverse mass distributions for $gg\rightarrow H
\rightarrow ZZ\rightarrow\ell^+\ell^-\nu\bar{\nu}$ (dashed),
$q\bar{q}\rightarrow
ZZ\rightarrow\ell^+\ell^-\nu\bar{\nu}$ (solid) and $q\bar{q}\rightarrow Z g
\rightarrow \ell^+\ell^- g$, with crossed diagrams, (dot-dashed)
for $pp$ collisions
at $\sqrt{S}=14$ TeV. Here, the lepton rapidity is $|y_\ell|<3$ and the
final state parton rapidity is $|y_p|>3$ to mimic a missing jet.
}
\end{center}

The decay distributions of the lepton from $Z_L$ and $Z_T$ decays, in
the $Z$ rest frame, are
\begin{equation}
\phi_L(z)={3\over 4}(1-z^2)\ \ \
\phi_T(z)={3\over 8} (1+z^2)
\end{equation}
where $z=\cos\theta$ for $\theta$, the angle between the lepton and the
axis defined by the $Z$ momentum in the parton center of mass frame.
The shape of the decay distribution as well as the average value of $|z|$
(for longitudinally polarized $Z$'s, a value of 3/8, and for transversely
polarized $Z$'s, a value of 9/16) characterize the production mechanism.
These decay distributions for the gold plated modes have been discussed
elsewhere.$^5$ For the Higgs decay into $ZZ$ where one $Z$ decays into
neutrinos, one loses the information required to reconstruct the
parton center of mass.

Note that
in the parton center of mass frame
($pcm$), $z=-2 p_\ell^{pcm}\cdot \epsilon_L^{pcm}/M_Z$,
where we have a four-vector dot product which involves the
longitudinal polarization vector $\epsilon_L^{pcm}$. The parton center of
mass frame differs from the hadron center of mass frame by a boost,
however, the Lorentz boost of $\epsilon_L^{pcm}$ converts the four vector
into a linear combination of longitudinal and transverse polarization vectors.
If the Higgs is heavy enough, the boost will not be large.
To approximate $|z|$ in the hadron center of mass, we introduce the variable
$z^*$ where $z^*=2| p_\ell\cdot\epsilon_L|/M_Z$. Here, all of the vectors
are in the hadron center of mass, and $\epsilon_L= (|\vec{p}_Z|/M_Z,
E_z\vec{p}_Z/(|\vec{p}_Z| M_Z))$
is the longitudinal polarization vector determined
from the reconstructed $Z$ momentum. In the results presented below,
we plot the average value of $z^*$ as a function of transverse mass.

\section{Results}

We now show our results for several values of the Higgs mass. In Fig. 2,
we show $<z^*>$ as a function of $m_T$ for $m_H=600$ GeV, where the
input top quark mass is taken at $m_t=175$ GeV. The triangles show the
$<z^*>$ value for the combination of
$ZZ$ continuum production and the QCD
missing jet background, and the squares are for the gluon fusion Higgs
signal.  For each of the separate cross sections $\sigma_i,\ i=c$ (continuum),
$m$ (missing jet background) and $h$ (Higgs signal), the overall
$<z^*>$ value is determined by
\begin{equation}
<z^*>={ <z^*>_h\sigma_h+\kappa(<z^*>_m\sigma_m+<z^*>_c\sigma_c)\over
\sigma_h+\kappa (\sigma_m+\sigma_c)}
\end{equation}
with $\kappa = 1$. In order to estimate the error on the theoretical
prediction for $<z^*>$, we also take $\kappa=1\pm 0.3$. This error
from the normalization uncertainty from non-Higgs cross sections is
combined in quadrature with a statistical error. The statistical error
is estimated by $<z^*>/\sqrt{N_{evt}}$, where the number of events assumes
an integrated luminosity of 10$^5$ pb$^{-1}$. The combined error is
the outer error bar, with only the statistical error indicated by
the inner vertical error bar on the figures. We take a transverse mass
bin width of 50 GeV, except for the two last bins, where the widths are
100 GeV and 250 GeV, respectively, to include a reasonable number of events.
The statistical error dominates the overall error.
This figure and Figs. 3$a$ and 3$b$ below use $y_\ell^c=y_p^c=3.0$.
Figure 3$a$ shows $<z^*>$ versus $m_T$ for $m_H=400$ GeV, while Fig. $3b$ has
$m_H=800$ GeV. In all of these figures, only one charged lepton family is
included in the rate.

\vspace*{5.0cm}
\begin{center}
{\small Fig. 2. The value of $<z^*>$ as a function of $m_T$
for $m_H=600$ GeV and $m_t=175$ GeV. Rapidity cuts
applied are $y_\ell^c=y_p^c=3.0$.
The symbols are described in the
text.}
\end{center}

At the LHC, a more realistic set of rapidity cuts is to include charged
leptons in a more central region, with $y_\ell ^c=2.5$, and to put
the ``missing jet'' cut at $y_p^c=4$ for the parton rapidity.
With these cuts, the signal event rate is not reduced very much, but
the QCD missing jet background is greatly reduced. Figures 4$a$ and 4$b$
show $<z^*>$ versus $m_T$ for $m_H=400$ GeV and $m_H=800$ GeV, respectively,
with these new rapidity cuts. Again, the error bars indicate
the combined normalization
and statistical error bars as described above.

Figures 2-4 indicate that the value of $<z^*>$ will be a useful
tool to characterize high transverse mass data at the LHC.
With an integrated luminosity of $10^5$ pb$^{-1}$, Higgs masses up
to 800 GeV are accessible using this method.
%Even with the lower integrated luminosity, Higgs bosons with
%masses between 400 GeV and 800 GeV should be observable using the polarization
%technique.

\vspace*{5.0cm}
\begin{center}
{\small Fig. 3. Value of $<z^*>$ for $m_T$ bins with
$a$) $m_H=400$ GeV and $b$) $m_H=800$ GeV, with $y_\ell^c=y_p^c=3$.}
\end{center}

\vspace*{5.0cm}
\begin{center}
{\small Fig. 4. As in Fig. 3, but with $y_\ell^c=2.5$ and $y_p^c=4$.}
\end{center}

\bibliographystyle{unsrt}

\end{document}